\documentclass[twocolumn,showpacs,showkeys,preprintnumbers]{revtex4-1}
\usepackage{amssymb}

\usepackage{amsfonts}

\usepackage{latexsym}

\usepackage{dcolumn}

\usepackage{amsmath}

\usepackage{graphicx}

\usepackage{psfrag,pstricks}

\begin{document}

\title{Phase noise and squeezing spectra of the output field of an optical cavity containing an interacting Bose-Einstein condensate}

\author{A. Dalafi$^{1}$ }
\email{adalafi@yahoo.co.uk}
\author{M. H. Naderi$^{2}$}

\affiliation{$^{1}$ Laser and Plasma Research Institute, Shahid Beheshti University, Tehran 1983969411, Iran\\
$^{2}$Quantum Optics Group, Department of Physics, Faculty of Science, University of Isfahan, Hezar Jerib, 81746-73441, Isfahan, Iran}

\date{\today}

\begin{abstract}
We present a theoretical study of the phase noise, intensity and quadrature squeezing power spectra of the transmitted field of a driven optical cavity containing an interacting one-dimensional Bose-Einstein condensate. We show how the pattern of the output power spectrum of the cavity changes due to the nonlinear effect of atomic collisions. Furthermore, it is shown that due to a one-to-one correspondence between the splitting of the peaks in the phase noise power spectrum of the cavity output field and the \textit{s}-wave scattering frequency of the atom-atom interaction, one can measure the strength of interatomic interaction. Besides, we show how the atomic collisions affect the squeezing behavior of the output field.
\end{abstract}

\maketitle

\section{Introduction}
Nowadays, optical cavities containing Bose-Einstein condensates (BECs), the so-called hybrid systems, have been considered as good candidates for investigation of atom-photon interaction in the regime where their quantum mechanical properties are manifested in the same level \citep{Maschler2008,dom JOSA}. Besides, phenomena typical of solid-state physics like the formation of energy bands \cite{Bha Opt. Commun} and Bloch oscillations \cite{Prasanna13} are clearly observable in these kinds of systems.

In these hybrid systems, an effective optomechanical coupling \cite{op1,op2,op3,op4} arises due to the dispersive interaction of the BEC matter field with the optical field of the cavity where the fluctuations of the atomic field of the BEC (the Bogoliubov mode) plays the role of the vibrational mode of the moving mirror in an optomechanical cavity \cite{Kanamoto, Nagy Ritsch 2009}. Furthermore, due to the high density of atoms, the nonlinear effect of atom-atom interaction plays a very important role in systems consisting of BEC \cite{Morsch, Szirmai 2010}.

One of the interesting features of the optomechanical systems is the possibility of the normal-mode splitting (NMS), i.e., the coupling of two degenerate modes with energy exchange taking place on a time scale faster than the decoherence of
each mode \cite{Dobr}. Optomechanical NMS may be taken into account in experiments that seek to demonstrate ground-state cooling which is possible in the resolved sideband regime, where the frequency of the mechanical mirror is much larger than the cavity decay rate \cite{Rae,Mar,Schliesser}. The more optomechanical coupling strength, the better the occurrence of ground-state cooling and NMS.

In recent years, it has been shown that using the nonlinear media such as an optical parametric amplifier (OPA) \cite{Hu1,Hu2}, an optical Kerr medium \cite{Kumar} or a combination of both \cite{Shahidani1,Shahidani2} lead to the enhancement of the radiation pressure-induced coupling.

On the other hand, an interacting BEC inside an optical cavity behaves like a nonlinear medium which possesses the nonlinear properties of both a Kerr medium \cite{dalafi1} and an OPA \cite{dalafi3}. In our two previous papers \cite{dalafi1,dalafi3} we have explicitly shown that an interacting BEC behaves as a so-called atomic parametric amplifier, similar to an OPA, where the condensate and the Bogoliubov modes play, respectively, the roles of the pump field and the signal mode in the degenerate parametric amplifier and the \textit{s}-wave scattering frequency of atom-atom interaction plays the role of the nonlinear gain parameter. Besides, it has been shown that these nonlinearities may affect the optical bistability of the cavity \cite{dalafi2} and change the threshold of the quantum phase transition of the BEC \cite{dalafi4}.

The above-mentioned investigations have been mostly concentrated on the influences of the nonlinear effects on the atomic field of the BEC or on intracavity optical field. However, to the best of our knowledge, nowhere it has been addressed how the nonlinearity induced by the atomic collisions affects the power spectrum of the output field of the cavity which is measurable experimentally. Addressing this question is the main theme of the present work.   

In this paper we study the nonlinear effect of the atomic collisions on the output power spectrum of a driven optical cavity containing a trapped one-dimensional BEC. We show that due to a one-to-one correspondence between the splitting of the peaks in the phase noise power spectrum of the cavity output field and the \textit{s}-wave scattering frequency of the atom-atom interaction, one can measure the strength of atomic interactions. Besides, we show how the atomic collisions affect the squeezing behavior of the output field.

The paper is structured as follows. In Sec.II we describe the system under consideration and obtain the dynamical Heisenberg-Langevin equations for the system operators. In Sec. III we show the relation of the \textit{s}-wave scattering frequency and the splitting of the peaks of the phase noise power spectrum of the transmitted field. In Secs. IV and V we investigate the effect of atomic collisions on the intensity spectrum and the quadrature squeezing of the transmitted field, respectively. Finally, our conclusions are summarized in Sec. VI.

\section{Description of the system Hamiltonian and Dynamics}\label{secH}
\begin{figure}[ht]
\centering
\includegraphics[width=2.7in]{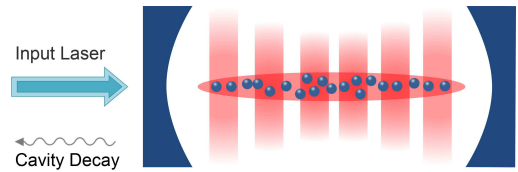}
\caption{
(Color online) A Bose-Einstein condensate trapped inside an optical cavity interacting with a single cavity mode. The cavity mode is driven by a laser at rate $\eta$ and the decay rate is $\kappa$.}
\label{fig:fig1}
\end{figure}

As depicted in Fig.\ref{fig:fig1}, we consider a system consisting of an optical cavity with length $L$ containing a BEC of $N$ two-level atoms with mass $m_{a}$ and transition frequency $\omega_{a}$ which is driven at rate $\eta$ through one of the end mirrors by a laser with frequency $\omega_{p}$, and wave number $k=\omega_{p}/c$. If the BEC is confined in a cylindrically symmetric trap with a transverse trapping frequency $\omega_{\mathrm{\perp}}$ and negligible longitudinal confinement along the $x$ direction \cite{Morsch}, then the dynamics can be described within an effective one-dimensional model by quantizing the atomic motional degree of freedom along the $x$ axis only.

In the dispersive regime, where the laser pump is detuned far above the atomic resonance ($\Delta_{a}=\omega_{p}-\omega_{a}>0$  exceeds the atomic linewidth $\gamma$ by orders of magnitude), the excited electronic state of the atoms can be adiabatically eliminated and spontaneous emission can be neglected \cite{Masch Ritch 2004}. In the frame rotating at the pump frequency, the many-body Hamiltonian reads
\begin{eqnarray}\label{H1}
&H&=-\hbar\Delta_{c} a^{\dagger} a+i\hbar\eta (a-a^{\dagger})+\int_{-\frac{L}{2}}^{\frac{L}{2}}\Psi^{\dagger}(x)\Big[\frac{-\hbar^{2}}{2m_{a}}\frac{d^{2}}{dx^{2}}\nonumber\\
&&+\hbar U_{0} \cos^2(kx) a^{\dagger} a+\frac{1}{2} U_{s}\Psi^{\dagger}(x)\Psi(x)\Big] \Psi(x) dx.
\end{eqnarray}
Here, $a$ is the annihilation operator of the optical field, $\Delta_{c}=\omega_{p}-\omega_{c}$ is the cavity-pump detuning, $U_{0}=g_{0}^{2}/\Delta_{a}$ is the optical lattice barrier height per photon which represents the atomic back action on the field, $g_{0}$ is the vacuum Rabi frequency, $U_{s}=\frac{4\pi\hbar^{2} a_{s}}{m_{a}w^{2}}$, $a_{s}$ is the two-body \textit{s}-wave scattering length \cite{Masch Ritch 2004,Dom JB}, and $ w $ is the waist of the optical potential.

In the regime, where $U_{0}\langle a^{\dagger}a\rangle\leq 10\omega_{R}$ ($\omega_{R}=\frac{\hslash k^{2}}{2m_{a}}$ is the recoil frequency of the condensate atoms), and under the Bogoliubov approximation \cite{Nagy Ritsch 2009}, the atomic field operator can be expanded as the following single-mode quantum field

\begin{equation}\label{opaf}
\Psi(x)=\sqrt{\frac{N}{L}}+\sqrt{\frac{2}{L}}\cos(2kx) c,
\end{equation}
where the first term is the condensate mode which is considered as a c-number and the operator $c$ in the second term is the annihilation operator of the Bogoliubov mode (matter-field fluctuations). Substituting this expansion into Eq.(\ref{H1}), the Hamiltonian of the system takes the form
\begin{eqnarray}\label{Hopa}
&H&=\hbar\delta_{c}a^{\dagger}a+i\hbar\eta (a-a^{\dagger})+\hbar\Omega_{c}c^{\dagger}c\nonumber\\
&&+\frac{1}{4}\hbar\omega_{sw}(c^{2}+c^{\dagger2})+\frac{\sqrt{2}}{2}\hbar\zeta a^{\dagger}a (c+c^{\dagger}).
\end{eqnarray}
In this equation, $\delta_{c}=-\Delta_{c}+\frac{1}{2}N U_{0}$ is the cavity Stark-shifted detuning and $ \Omega_{c}=4\omega_{R}+\omega_{sw} $ denotes the frequency of the Bogoliubov mode. The fourth term of the Hamiltonian
(\ref{Hopa}) corresponds to the atomic collisions in which $ \omega_{sw}=8\pi\hbar a_{s}N/m_{0}Lw^2 $ is the \textit{s}-wave scattering frequency. The last term describes the optomechanical coupling between the Bogoliubov mode and the radiation pressure of the optical field with the coupling constant $ \zeta=\frac{1}{2}\sqrt{N}U_{0} $. In Ref.\cite{Nagy2013} it has been shown that by a suitable Bogoliubov transformation the Hamiltonian of Eq.(\ref{Hopa}) can be transformed to an ordinary optomechanical Hamiltonian with a modified optomechanical coupling.

To study the dynamics of the system, we use the Heisenberg equations considering the noise sources and dissipation of both the optical and the matter fields to obtain the following nonlinear quantum Langevin equations (QLEs) of motion
\begin{subequations}\label{NQL}
\begin{eqnarray}\label{subNQL}
\dot{a}&=&-(i\delta_{c}+\kappa)a-\frac{i}{\sqrt{2}}\zeta a(c+c^{\dagger})-\eta+\sqrt{2\kappa} \delta a_{in},\\
\dot{c}&=&-(i\Omega_{c}+\gamma)c-\frac{i}{2}\omega_{sw}c^{\dagger}-\frac{i}{\sqrt{2}}\zeta a^{\dagger}a+\sqrt{2\gamma} \delta c_{in}.
\end{eqnarray}
\end{subequations}
Here, $\kappa$ and $\gamma$ characterize the dissipation of the cavity field and collective density excitations of the BEC, respectively. The cavity-field quantum vacuum fluctuation $\delta a_{in}(t)$ satisfies the Markovian correlation functions, i.e., $\langle\delta a_{in}(t)\delta a_{in}^{\dagger}(t^{\prime})\rangle=(n_{ph}+1)\delta(t-t^{\prime})$, $\langle\delta a_{in}^{\dagger}(t)\delta a_{in}(t^{\prime})\rangle=n_{ph}\delta(t-t^{\prime})$ with the average thermal photon number $n_{ph}$ which is nearly zero at optical frequencies \cite{Gardiner}. Besides, $\delta c_{in}(t)$ is the thermal noise input for the Bogoliubov mode of the BEC which also satisfies the same Markovian correlation functions as those of the optical noise. The noise sources are assumed uncorrelated for the different modes of both the matter and light fields.

The set of nonlinear equations (\ref{NQL}) can be linearized by decomposing each operator as the sum of its steady-state value and a small fluctuation. In this way, we can substitute $ a=\alpha+\delta a $ and $ c=\sqrt{N}\beta +\delta c $ into Eq.(\ref{NQL}) to obtain a set of nonlinear algebraic equations for the mean-field values and another set of linear ordinary differential equations for the quantum fluctuations. The steady-state mean-field values are obtained as follows
\begin{subequations}\label{ss}
\begin{eqnarray}
\alpha&=&\frac{\eta}{\sqrt{\Delta_{d}^2+\kappa^2}},\\
\beta&=&\frac{\sqrt{2}}{4}\frac{U_{0}\alpha^2}{\sqrt{(\Omega_{c}+\frac{1}{2}\omega_{sw})^{2}+\gamma^2}},
  \end{eqnarray}
\end{subequations}
where $ \Delta_{d}=-\delta_{c}-\sqrt{2N}\zeta\beta $ is the effective detuning and we have assumed $ \alpha $ and $ \beta $ are real numbers\cite{Konya Domokos 2011}. Furthermore, the linearized QLEs are obtained from the following equations
\begin{subequations}\label{flc ac}
\begin{eqnarray}
\delta\dot{a}&=&(i\Delta_{d}-\kappa)\delta a-\frac{i}{2}G(\delta c+\delta c^{\dagger})+\sqrt{2\kappa}\delta a_{in},\label{ac1}\\
\delta\dot{c}&=&-(i\Omega_{c}+\gamma)\delta c-\frac{i}{2}G(\delta a+\delta a^{\dagger})-\frac{i}{2}\omega_{sw}\delta c^{\dagger}+\sqrt{2\gamma}\delta c_{in},\label{ac2}\nonumber\\
\end{eqnarray}
\end{subequations}
where $ G=\sqrt{2}\zeta\alpha $ is the enhanced optomechanical coupling rate for the linearized regime. Introducing the optical and matter-field quadratures
\begin{subequations}\label{quadratures}
\begin{eqnarray}
\delta X_{a}&=&\frac{1}{\sqrt{2}}(\delta a+\delta a^{\dagger}),  \delta P_{a}=\frac{1}{\sqrt{2}i}(\delta a-\delta a^{\dagger}),\\
\delta X_{c}&=&\frac{1}{\sqrt{2}}(\delta c+\delta c^{\dagger}),  \delta P_{c}=\frac{1}{\sqrt{2}i}(\delta c-\delta c^{\dagger}),
\end{eqnarray}
\end{subequations}
and the input noise quadratures 
\begin{subequations}\label{noisequads}
\begin{eqnarray}
\delta X_{a}^{(in)}&=&\frac{1}{\sqrt{2}}(\delta a_{in}+\delta a^{\dagger}_{in}),  \delta P_{a}^{(in)}=\frac{1}{\sqrt{2}i}(\delta a_{in}-\delta a^{\dagger}_{in}),\nonumber\\
\delta X_{c}^{(in)}&=&\frac{1}{\sqrt{2}}(\delta c_{in}+\delta c^{\dagger}_{in}),  \delta P_{c}^{(in)}=\frac{1}{\sqrt{2}i}(\delta c_{in}-\delta c^{\dagger}_{in}),\nonumber
\end{eqnarray}
\end{subequations}
the QLEs can be written in the compact matrix form
\begin{equation}\label{nM}
\dot{u}(t)=M u(t)+n(t),
\end{equation}
where $u=[\delta X_{a},\delta P_{a},\delta X_{c},\delta P_{c}]^{T}$ is the vector of continuous variable fluctuation operators and
$ n(t)=[\sqrt{2\kappa}\delta X_{a}^{(in)},\sqrt{2\kappa}\delta P_{a}^{(in)},\sqrt{2\gamma}\delta X_{c}^{(in)},\sqrt{2\gamma}\delta P_{a}^{(in)}]^{T} $ is the corresponding vector of noises. The $4\times4$ matrix $M$ is the drift matrix given by
\begin{equation}
M=\left(\begin{array}{cccc}
-\kappa & -\Delta_{d} & 0 & 0 \\
   \Delta_{d} & -\kappa &-G &0 \\
    0 & 0 & -\gamma & \Omega^{(-)}_{c} \\
    -G & 0 & -\Omega^{(+)}_{c} & -\gamma\\
    \end{array}\right),
\label{M}
\end{equation}
where $ \Omega_{c}^{(\pm)}=\Omega_{c}\pm\frac{1}{2}\omega_{sw} $. The solutions to Eq.(\ref{nM}) are stable only if all the eigenvalues of the matrix $ M $ have negative real parts. The stability conditions can be obtained, for example, by using the Routh-Hurwitz criteria \cite{RH}.

\section{Phase noise spectrum}
The output power spectrum of the phase quadrature of the optical field of the cavity is an interesting quantity which is experimentally measurable by the homodyne measurement of the light reflected by the cavity \cite{Gio}. It is defined by the following equation in the frequency space
\begin{eqnarray}
S_{P}(\omega)&=&\frac{1}{4\pi}\int d\omega^{\prime} e^{i(\omega+\omega^{\prime})t}\big\langle\delta P_{a}^{(out)}(\omega)\delta P_{a}^{(out)}(\omega^{\prime})\nonumber\\
&&+\delta P_{a}^{(out)}(\omega^{\prime})\delta P_{a}^{(out)}(\omega)\big\rangle.
\end{eqnarray}
Using the input-output relation $ \delta P_{a}^{(out)}(\omega)=\sqrt{2\kappa}\delta P_{a}(\omega)-\delta P_{a}^{(in)}(\omega) $ \cite{Walls} the phase noise power spectrum can be written as
\begin{eqnarray}\label{Spa}
S_{P}(\omega)&=&\frac{1}{2}+\frac{\kappa}{2\pi}\int d\omega^{\prime} e^{i(\omega+\omega^{\prime})t}\nonumber\\
&&\times\big\langle\delta P_{a}(\omega)\delta P_{a}(\omega^{\prime})+\delta P_{a}(\omega^{\prime})\delta P_{a}(\omega)\big\rangle,
\end{eqnarray}
where we have assumed that the mean thermal photon number, $ n_{ph} $, is zero. In order to calculate this power spectrum one needs to solve the time-domain equation of motion by Fourier transforming it into the frequency domain. The Fourier transform of an operator like $ \delta F(t) $ is
\begin{equation}\label{FT}
\delta F(t)=\frac{1}{2\pi}\int_{-\infty}^{+\infty}\delta F(\omega) e^{i\omega t} d\omega.
\end{equation}
In this way the set of differential equations (\ref{nM}) is transformed into a set of algebraic equations in the frequency space. Solving them for the phase quadrature of the optical field yields
\begin{eqnarray}\label{Pa}
\delta P_{a}(\omega)&=&\chi (\omega)\Big[f_{1}(\omega)\delta X_{a}^{(in)}(\omega)+f_{2}(\omega)\delta P_{a}^{(in)}(\omega)\nonumber\\
&&+f_{3}(\omega)\delta X_{c}^{(in)}(\omega)+f_{4}(\omega)\delta P_{c}^{(in)}(\omega)\Big],
\end{eqnarray}
where the coefficients $ f_{i} $ are given by
\begin{subequations}\label{f}
\begin{eqnarray}
f_{1}(\omega)&=&\frac{\sqrt{2\kappa}\big[\Delta_{d}\big((\gamma+i\omega)^2+\omega_{m}^2\big)+G^2\Omega_{c}^{(-)}\big]}{\Delta_{d}^2+(\kappa+i\omega)^2},\\
f_{2}(\omega)&=&\frac{\sqrt{2\kappa}(\kappa+i\omega)\big[(\gamma+i\omega)^2+\omega_{m}^2\big]}{\Delta_{d}^2+(\kappa+i\omega)^2},\\
f_{3}(\omega)&=&-\frac{G\sqrt{2\gamma}(\gamma+i\omega)(\kappa+i\omega)}{\Delta_{d}^2+(\kappa+i\omega)^2},\\
f_{4}(\omega)&=&-\frac{G\sqrt{2\gamma}\Omega_{c}^{(-)}(\kappa+i\omega)}{\Delta_{d}^2+(\kappa+i\omega)^2}.
\end{eqnarray}
\end{subequations}
In these equations
\begin{equation}\label{wm}
\omega_{m}=\sqrt{\Omega_{c}^{(+)}\Omega_{c}^{(-)}}=\sqrt{(4\omega_{R}+\omega_{sw}/2)(4\omega_{R}+3\omega_{sw}/2)}
\end{equation}
is the oscillation frequency of the mode $ (\delta X_{c},\delta P_{c}) $ which behaves just like the mechanical oscillator in an optomechanical cavity. Furthermore, the susceptibility $ \chi(\omega) $ in Eq.(\ref{Pa}) is given by
\begin{equation}\label{chi}
\chi=\Big[(\gamma+i\omega)^2+\omega_{m}^2+\frac{G^2\Delta_{d}\Omega_{c}^{(-)}}{\Delta_{d}^2+(\kappa+i\omega)^2}\Big]^{-1}.
\end{equation}

In order to obtain the power spectrum we need the correlation functions of the noise sources in the frequency domain. For the cavity-field quantum vacuum fluctuations we have
\begin{subequations}\label{cor a(w)}
\begin{eqnarray}
\langle\delta a_{in}(\omega)\delta a_{in}^{\dagger}(\omega^{\prime})\rangle&=&2\pi\delta(\omega+\omega^{\prime}),\\
\langle\delta a_{in}^{\dagger}(\omega)\delta a_{in}(\omega^{\prime})\rangle&=&0,\\
\langle\delta a_{in}(\omega)\delta a_{in}(\omega^{\prime})\rangle&=&0,\\
\langle\delta a_{in}^{\dagger}(\omega)\delta a_{in}^{\dagger}(\omega^{\prime})\rangle&=&0,
\end{eqnarray}
\end{subequations}
and also for their corresponding quadrature vacuum fluctuation
\begin{subequations}\label{cor XP(w)}
\begin{eqnarray}
\langle\delta X^{(in)}_{a}(\omega)\delta X^{(in)}_{a}(\omega^{\prime})\rangle&=&\pi\delta(\omega+\omega^{\prime}),\\
\langle\delta P^{(in)}_{a}(\omega)\delta P^{(in)}_{a}(\omega^{\prime})\rangle&=&\pi\delta(\omega+\omega^{\prime}),\\
\langle\delta X^{(in)}_{a}(\omega)\delta P^{(in)}_{a}(\omega^{\prime})\rangle&=&i\pi\delta(\omega+\omega^{\prime}),\\
\langle\delta P^{(in)}_{a}(\omega)\delta X^{(in)}_{a}(\omega^{\prime})\rangle&=&-i\pi\delta(\omega+\omega^{\prime}).
\end{eqnarray}
\end{subequations}
For the input noise of the Bogoliubov mode of the BEC, there are similar relations like Eqs.(\ref{cor a(w)},\ref{cor XP(w)}). Substituting Eq.(\ref{Pa}) into the power spectrum of Eq.(\ref{Spa}) and using Eqs.(\ref{cor a(w)},\ref{cor XP(w)}), the output power spectrum of the phase quadrature of the optical field of the cavity will be obtained as
\begin{eqnarray}\label{Sp}
S_{P}(\omega)&=&\frac{1}{2}+\kappa \vert\chi(\omega)\vert^2\Big(\vert f_{1}(\omega)\vert^2+\vert f_{2}(\omega)\vert^2\nonumber\\
&&+\vert f_{3}(\omega)\vert^2+\vert f_{4}(\omega)\vert^2\Big).
\end{eqnarray}

Based on Eqs.(\ref{f},\ref{chi}), there is a common factor $ \vert D(\omega)\vert^2 $ in the denominator of all terms of Eq.(\ref{Sp}) where $ D(\omega) $ is given by
\begin{equation}\label{D}
D(\omega)=\big[(\gamma+i\omega)^2+\omega_{m}^2\big]\big[\Delta_{d}^2+(\kappa+i\omega)^2\big]+G^2\Delta_{d}\Omega_{c}^{(-)}.
\end{equation}
Therefore, the peaks of the spectrum occur at the roots of Eq.(\ref{D}). If the damping rates $ \gamma $ and $ \kappa $ are small in comparison to $ \omega_{m} $ and $ \Delta_{d} $ then the positive roots of Eq.(\ref{D}) can be approximated with the following frequencies
\begin{equation}\label{D roots}
\omega_{\pm}\simeq\sqrt{\frac{1}{2}(\omega_{m}^2+\Delta_{d}^2)\pm\frac{1}{2}\sqrt{(\omega_{m}^2-\Delta_{d}^2)^2-4G^2\Delta_{d}\Omega_{c}^{(-)}}}.
\end{equation}

As is seen from Eq.(\ref{D roots}), in the absence of optomechanical coupling where $ G=0 $, the spectrum has two peaks at $ \omega=\pm\omega_{m} $ corresponding to the frequency of the so-called mechanical mode of the BEC, and also it has two other peaks at $ \omega=\pm\Delta_d $ corresponding to the optical effective frequency. However, the optomechanical coupling changes the positions of these maxima around $ \omega_{m} $ and $ \Delta_{d} $ according to Eq.(\ref{D roots}). The amount of splitting between the two modes can be analytically obtained as $ \Delta\omega_{\pm}=\omega_{+}-\omega_{-} $ from Eq.(\ref{D roots}) and can also be calculated numerically by taking the difference between the imaginary parts of the eigenvalues of the matrix $ M $.

\begin{figure}[ht]
\centering
\includegraphics[width=2.8 in]{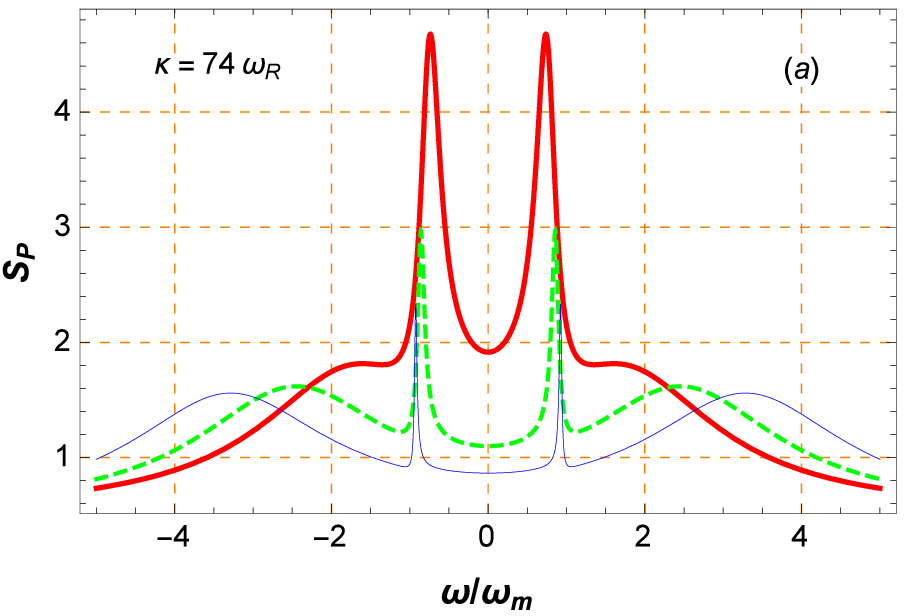}
\includegraphics[width=2.8 in]{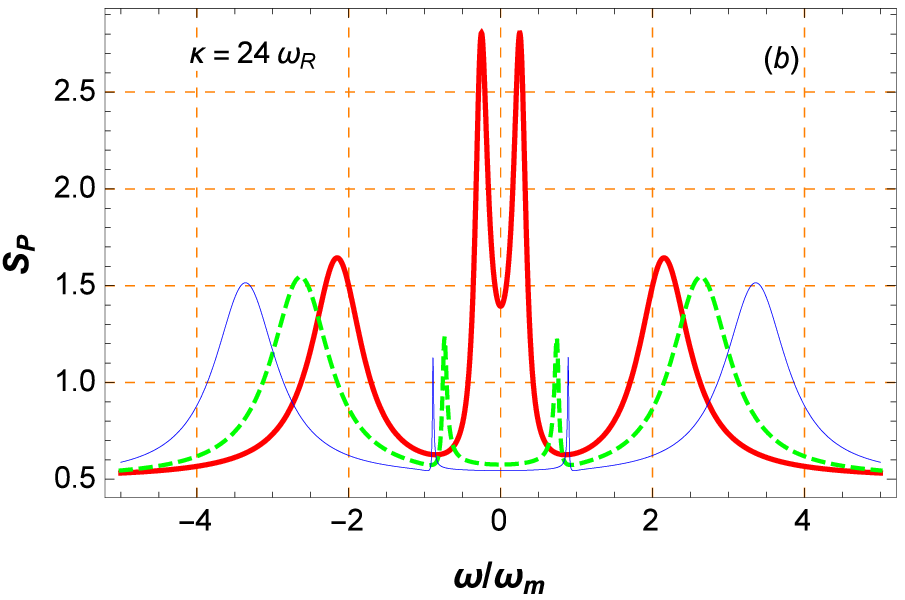}
\caption{
(Color online)The phase noise spectrum of the output field of the cavity versus the normalized frequency $ \omega/\omega_{m} $ for three different values of the effective detuning $ \delta_{c}=\omega_{m} $ (red thick line), $ \delta_{c}=2\omega_{m} $ (green dashed line), and $ \delta_{c}=3\omega_{m} $ (blue thin line) for two different values of the cavity damping rate: (a)$ \kappa=74\omega_{R} $ and (b) $ \kappa=24\omega_{R} $ . The parameters are $ L=187 \mu$m, $ \lambda=780 $nm, $ \eta=81\omega_{R} $, $ \gamma=0.001\kappa $, and $ \omega_{sw}=50\omega_{R} $.}
\label{fig:fig2}
\end{figure}

\subsection{Effect of detuning on the spectrum}
Now we can analyse our results based on the experimentally feasible parameters given in \cite{Ritter Appl. Phys. B, Brenn Science},i.e., we assume there are $ N=10^5 $ Rb atoms inside an optical cavity of length $ L=187 \mu$m with bare frequency $ \omega_{c}=2.41494\times 10^{15} $Hz corresponding to a wavelength of $ \lambda=780 $nm. The mode has a waist radius of $ 25 \mu$m and is coherently driven at amplitude $ \eta $ through one of the cavity mirrors with a pump laser at frequency $ \omega_{p} $. The atomic $ D_{2} $ transition corresponding to the atomic transition frequency $ \omega_{a}=2.41419\times 10^{15} $Hz couples to the mentioned mode of the cavity. The atom-field coupling strength $ g_{0}=2\pi\times 14.1 $MHz and the recoil frequency of the atoms is $ \omega_{R}=23.7 $KHz.

\begin{figure}[ht]
\centering
\includegraphics[width=2.8in]{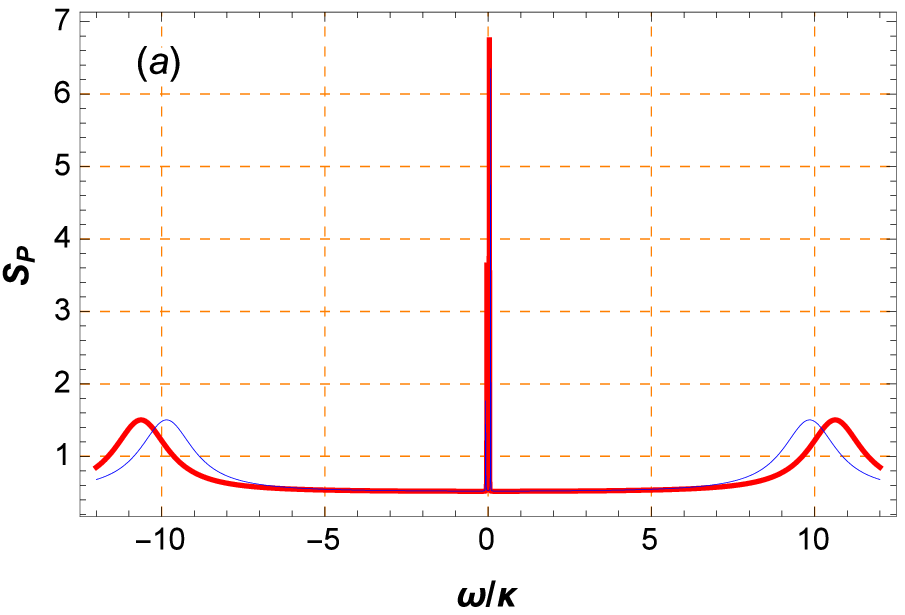}
\includegraphics[width=2.8in]{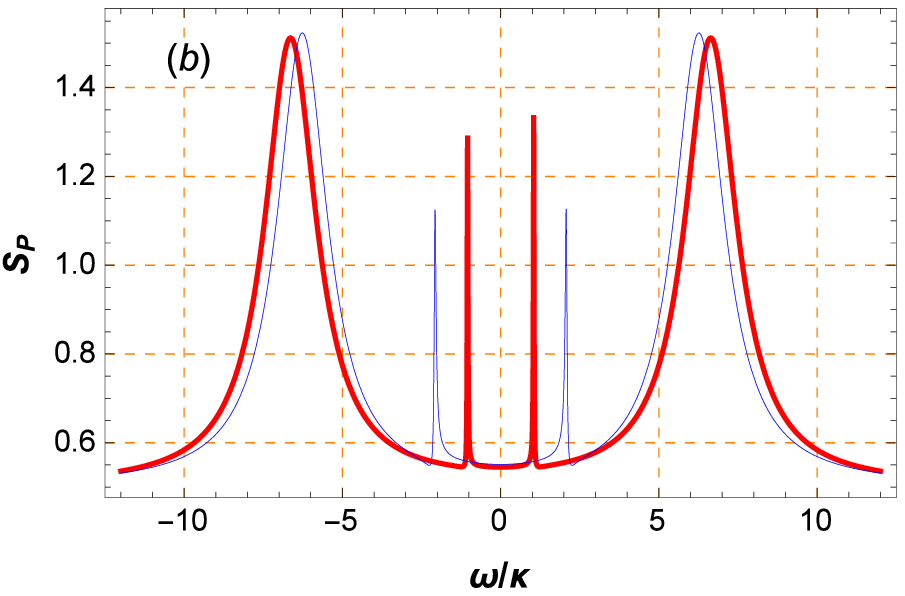}
\caption{
(Color online) The phase noise spectrum of the output field of the cavity versus the normalized frequency $ \omega/\kappa $ for a fixed value $ \Delta_{c}=0.994\Delta_{0} $,for four different \textit{s}-wave scattering frequency: (a) $ \omega_{sw}=0 $ (red thick line) and $ \omega_{sw}=1\omega_{R} $ (blue thin line) and (b)$ \omega_{sw}=30\omega_{R} $ (red thick line) and $ \omega_{sw}=60\omega_{R} $ (blue thin line). The parameters are $ L=187 \mu$m, $ \lambda=780 $nm, $ \eta=81\omega_{R} $, $ \kappa=24\omega_{R} $, and $ \gamma=0.001\kappa $,.}
\label{fig:fig3}
\end{figure}

In Fig.\ref{fig:fig2} we have plotted the phase noise spectrum of the output field of the cavity versus the normalized frequency $ \omega/\omega_{m} $ for three different values of the effective detuning $ \delta_{c}=\omega_{m} $ (red thick line), $ \delta_{c}=2\omega_{m} $ (green dashed line), and $ \delta_{c}=3\omega_{m} $ (blue thin line) for two different values of the cavity damping rate: $ \kappa=74\omega_{R} $ [Fig.\ref{fig:fig2}(a)] and $ \kappa=24\omega_{R} $ [Fig.\ref{fig:fig2}(b)]. Here, we have assumed that the cavity is pumped at rate $ \eta=81\omega_{R} $ and the \textit{s}-wave scattering frequency of atomic interactions is $ \omega_{sw}=50\omega_{R} $. For this value of $ \omega_{sw} $, the so-called mechanical frequency of the Bogoliubov mode of the BEC is obtained from Eq.(\ref{wm}) as $ \omega_{m}=47.86\omega_{R} $. Besides, the damping rate of the collective density excitations of the BEC has been considered to be $ \gamma=0.001\kappa $.

As is seen from Fig.\ref{fig:fig2}, each curve has four peaks corresponding to the positive and negative frequencies of the mechanical and optical modes. For a specified value of \textit{s}-wave scattering frequency, one can use the effective detuning $ \delta_{c} $ as an experimentally controllable parameter (through $ \omega_{p} $) to manipulate the amount of the splitting between the two modes. For example in Fig.\ref{fig:fig2} where $ \omega_{sw}=50\omega_{R} $ by increasing $ \delta_{c} $ from $ \omega_{m} $ to $ 3\omega_{m} $ the splitting of the modes increases. It is because of the fact that the effective frequency of the optical mode,i.e., $ \Delta_{d} $ is shifted by changing $ \delta_{c} $. As is seen from Fig.\ref{fig:fig2}, increasing the effective detununig causes the first peak (corresponding to the so-called mechanical mode) gets near to $ \omega_{m} $ from the left and the second one (corresponding to the optical mode) is shifted to higher frequency according to Eq.(\ref{D roots}). 

The interesting point is that the amount of splitting between the peaks in the phase noise spectrum is considerable even when the system is not in the strong coupling regime where $ G>>\kappa $. In Fig.\ref{fig:fig2}(a) where $ G\sim\kappa $ the splitting between the peaks are considerable specially for larger values of $ \delta_{c} $. However, in Fig.\ref{fig:fig2}(b)where $ G>\kappa $ the splittings are much more considerable. Similar effects can be observed in optical cavities containing a quantum well \cite{Eleuch2011,Eleuch2012}.

\subsection{Effect of atomic collisions on the spectrum}
It is well known that a BEC inside an optical cavity acts as a Kerr medium that shifts the empty cavity resonance as much as $ \Delta_{0}=\frac{1}{2}NU_{0} $ \cite{Ritter Appl. Phys. B, dalafi1}. Since we are interested in the spectrum of the transmitted field in the regime where the system is stable we should study the behavior of the system for detunings $ \Delta_{c}\lesssim \Delta_{0} $. Based on our numerical calculations if we fix $ \Delta_{c}=0.994\Delta_{0} $ the system is stable for a wide range of $ \omega_{sw}\geqslant 0 $.

By fixing $ \Delta_{c}=0.994\Delta_{0} $, we have plotted in Fig.\ref{fig:fig3} the phase noise spectrum versus the normalized frequency $ \omega/\kappa $ for four different values of the \textit{s}-wave scattering frequency: $ \omega_{sw}=0 $ [Fig.\ref{fig:fig3}(a), red thick line], $ \omega_{sw}=1\omega_{R} $ [Fig.\ref{fig:fig3}(a), blue thin line], $ \omega_{sw}=30\omega_{R} $ [Fig.\ref{fig:fig3}(b), red thick line], and $ \omega_{sw}=60\omega_{R} $ [Fig.\ref{fig:fig3}(b), blue thin line]. As is seen form this figure, increasing the \textit{s}-wave scattering frequency makes the two peaks (corresponding to the optical and mechanical modes) get nearer to each other while the linewidth of the mechanical mode is increased and that of the optical mode is decreased.

\begin{figure}[ht]
\centering
\includegraphics[width=2.8in]{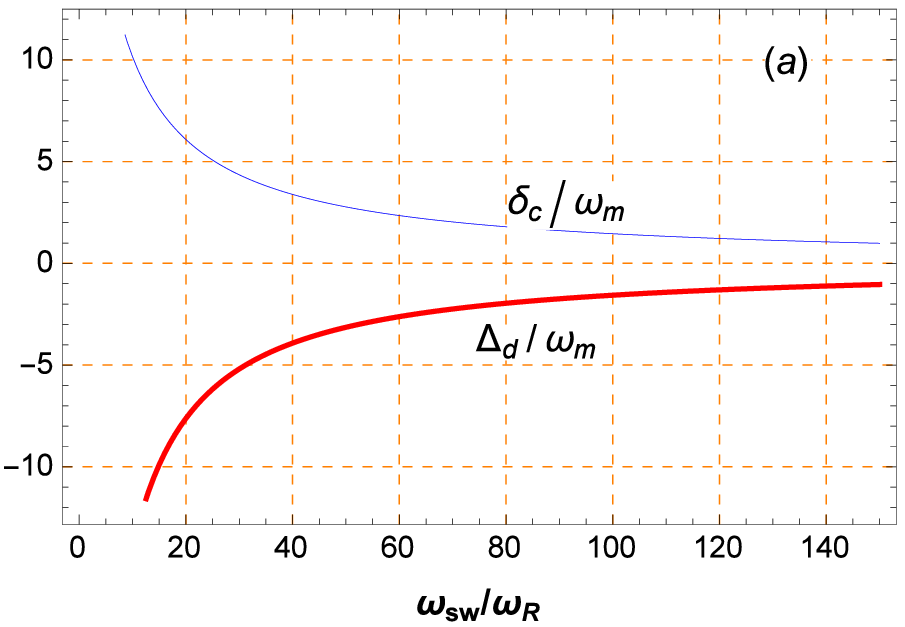}
\includegraphics[width=2.8in]{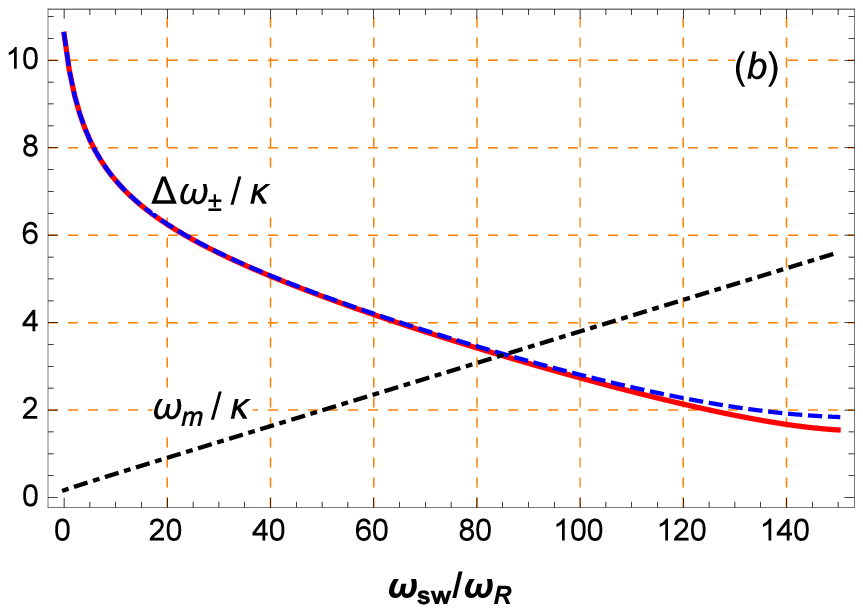}
\caption{
(Color online) (a)the normalized cavity Stark-shifted detuning $ \delta_{c}/\omega_{m} $ (blue thin line) and the normalized effective frequency of the optical mode $ \Delta_{d}/\omega_{m} $ (red thick line) versus the normalized \textit{s}-wave scattering frequency $ \omega_{sw}/\omega_{R} $ and (b) the normalized frequency splitting between the optical and mechanical modes $ \Delta\omega_{\pm}/\kappa $ [numerical (red solid line) and analytic (blue dashed line)] and the normalized mechanical frequency of the BEC (black dashed-dotted line) versus the normalized \textit{s}-wave scattering frequency $ \omega_{sw}/\omega_{R} $. Other parameters are the same as those in Fig.\ref{fig:fig3}.}
\label{fig:fig4}
\end{figure}

In order to see more clearly how the atom-atom interaction affects the pattern of output power spectrum, in Fig.\ref{fig:fig4} we have plotted the normalized cavity Stark-shifted detuning $ \delta_{c}/\omega_{m} $ [Fig.\ref{fig:fig4}(a), blue thin line] and the normalized effective frequency of the optical mode $ \Delta_{d}/\omega_{m} $ [Fig.\ref{fig:fig4}(a), red thick line] versus the normalized \textit{s}-wave scattering frequency $ \omega_{sw}/\omega_{R} $. Besides, we have plotted the normalized frequency splitting between the optical and mechanical modes $ \Delta\omega_{\pm}/\kappa $ [Fig.\ref{fig:fig4}(b), red solid line (numerical) and blue dashed line (analytic)] as well as the normalized mechanical frequency of the BEC [Fig.\ref{fig:fig4}(b), black dashed-dotted line] versus the normalized \textit{s}-wave scattering frequency $ \omega_{sw}/\omega_{R} $.

As is seen from Fig.\ref{fig:fig4}, for small values of the \textit{s}-wave scattering frequency, the frequency of the mechanical mode is very small ($ \omega_{m}\approx 4\omega_{R} $) while that of the optical mode is very large ($ \vert\Delta_{d}\vert\approx 250\omega_{R} $). Increasing the \textit{s}-wave scattering frequency makes $ \omega_{m} $ increase while $ \vert\Delta_{d}\vert $ decrease so that at large values of \textit{s}-wave scattering frequency $ \Delta_{d}\rightarrow-\omega_{m} $ while $ \delta_{c}\rightarrow +\omega_{m} $ [Fig.\ref{fig:fig4}(a)]. Sine the positions of the peaks of power spectrum occur around the two frequencies $ \omega_{m} $ and $ \vert\Delta_{d}\vert\ $ in the positive frequency range, the larger the frequency $ \omega_{sw} $, the nearer the peaks of the power spectrum to each other, as is seen in Fig.\ref{fig:fig3}.

The splitting between these peaks, i.e., $ \Delta\omega_{\pm}=\omega_{+}-\omega_{-} $, can be calculated analytically through Eq.(\ref{D roots}) which has been shown by the blue dashed line in Fig.\ref{fig:fig4}(b). However, this splitting can also be calculated numerically through the difference between the positive imaginary parts of the eigenvalues of the matrix $ M $ given by Eq.(\ref{M}) which has been shown by the red solid line in Fig.\ref{fig:fig4}(b). As is seen, the results of numeric and analytic calculations have a very good coincidence.

The important result obtained from this investigation is that there is a one to one correspondence between the \textit{s}-wave scattering frequency of the atoms and the splitting between the two peaks of the output power spectrum as is seen from Fig.\ref{fig:fig4}. In this way by measuring the splitting between the output power spectrum peaks one can estimate the value of the \textit{s}-wave scattering frequency of the atoms. Besides, since $ \omega_{sw} $ is controllable through the transverse trapping frequency $ \omega_{\perp} $ \cite{Morsch}, the mechanical frequency of the BEC can be controlled through its dependence on the \textit{s}-wave scattering frequency.

In the following section we will show that the intensity spectrum of the transmitted field is not as suitable as the phase noise spectrum for measuring the \textit{s}-wave scattering frequency of the atoms because the peaks in the output intensity spectrum are not observable as clearly as those in the phase noise spectrum.

\section{Intensity spectrum of the output field}\label{Sec In-sq}

The intensity power spectrum of the cavity output field is defined as \cite{Eleuch2012}
\begin{equation}
S_{I}(\omega)=\int_{-\infty}^{+\infty}d\tau e^{-i\omega\tau}\langle \delta a^{\dagger}_{out}(t+\tau)\delta a_{out}(t)\rangle.
\end{equation}
Using the input-output relations
\begin{subequations}
\begin{eqnarray}
\delta a_{out}(t)&=&\sqrt{2\kappa}\delta a(t)-\delta a_{in}(t),\label{in-out1}\\
\delta a_{out}^{\dagger}(t)&=&\sqrt{2\kappa}\delta a^{\dagger}(t)-\delta a_{in}^{\dagger}(t),\label{in-out2}
\end{eqnarray}
\end{subequations}
the output intensity power spectrum will be obtained as
\begin{equation}
S_{I}(\omega)=2\kappa\int_{-\infty}^{+\infty}d\tau e^{-i\omega\tau}\langle \delta a^{\dagger}(t+\tau)\delta a(t)\rangle,
\end{equation}
where we have again assumed that the average of thermal photon number is zero. Now, using Eq.(\ref{FT}) to obtain the Fourier transforms of $ \delta a(t) $ and $ \delta a^{\dagger}(t) $ we can write the power spectrum in the frequency domain as
\begin{equation}
S_{I}(\omega)=2\kappa C_{a^{\dagger}a}(\omega),
\end{equation}
where the symmetrical function $ C_{a^{\dagger}a}(\omega) $ is defined as follows
\begin{eqnarray}\label{Caa+}
C_{a^{\dagger}a}(\omega)&=&\frac{1}{4\pi} \int_{-\infty}^{+\infty}d\omega^{\prime} e^{i(\omega+\omega^{\prime})t}\nonumber\\
&&\times\langle \delta a^{\dagger}(\omega)\delta a(\omega^{\prime})+\delta a^{\dagger}(\omega^{\prime})\delta a(\omega)\rangle.
\end{eqnarray}

In order to calculate this function, one needs to solve Eqs.(\ref{ac1}, \ref{ac2}) in the frequency domain to obtain $ \delta a(\omega) $. $ \delta a^{\dagger}(\omega) $ can be obtained from the expression for $ \delta a(\omega) $, using the relation $ \delta a^{\dagger}(\omega)=[\delta a(-\omega)]^{\dagger} $. So, we will have
\begin{subequations}\label{a(w)}
\begin{eqnarray}
\delta a(\omega)&=&\chi(\omega)\Big[g_{1}(\omega)\delta a_{in}(\omega)+g_{2}(\omega)\delta a_{in}^{\dagger}(\omega)\nonumber\\
&&+g_{3}(\omega)\delta X_{c}^{(in)}(\omega)+g_{4}(\omega)\delta P_{c}^{(in)}(\omega)\Big],\label{a(w)1}\\
\delta a^{\dagger}(\omega)&=&\chi^{\ast}(-\omega)\Big[g_{1}^{\ast}(-\omega)\delta a^{\dagger}_{in}(\omega)+g_{2}^{\ast}(-\omega)\delta a_{in}(\omega)\nonumber\\
&&+g_{3}^{\ast}(-\omega)\delta X_{c}^{(in)}(\omega)+g_{4}^{\ast}(-\omega)\delta P_{c}^{(in)}(\omega)\Big],\label{a(w)2}\nonumber\\
\end{eqnarray}
\end{subequations}
where the coefficients $ g_{i}(\omega) $ are given by
\begin{subequations}\label{g}
\begin{eqnarray}
g_{1}(\omega)&=&\frac{\sqrt{2\kappa}}{\kappa+i(\omega-\Delta_{d})}\Big[\chi^{-1}+i\frac{\Omega_{c}^{(-)}G^2/2}{\kappa+i(\omega-\Delta_{d})}\Big],\\
g_{2}(\omega)&=&\frac{iG^2\sqrt{\kappa}\Omega_{c}^{(-)}}{\sqrt{2}[\kappa+i(\omega-\Delta_{d})]^2},\\
g_{3}(\omega)&=&-\frac{iG\sqrt{\gamma}(\gamma+i\omega)}{[\kappa+i(\omega-\Delta_{d})]},\\
g_{4}(\omega)&=&-\frac{iG\sqrt{\gamma}\Omega_{c}^{(-)}}{[\kappa+i(\omega-\Delta_{d})]},
\end{eqnarray}
\end{subequations}
and $ \chi(\omega) $ has been defined by Eq.(\ref{chi}). Substituting Eqs.(\ref{a(w)1},\ref{a(w)2}) into Eq.(\ref{Caa+}) and using Eqs.(\ref{cor a(w)}) and (\ref{cor XP(w)}), the output intensity power spectrum will be obtained as follows
\begin{eqnarray}\label{SI}
S_{I}(\omega)&=&\frac{1}{2}\kappa \vert\chi(\omega)\vert^2\Big[2\vert g_{2}(\omega)\vert^2+2\vert g_{2}(-\omega)\vert^2\nonumber\\
&&+\vert g_{3}(\omega)\vert^2+\vert g_{3}(-\omega)\vert^2+\vert g_{4}(\omega)\vert^2+\vert g_{4}(-\omega)\vert^2\nonumber\\
&&+ig_{3}^{\ast}(\omega)g_{4}(\omega)-ig_{4}^{\ast}(\omega)g_{3}(\omega)\nonumber\\
&&+ig_{3}^{\ast}(-\omega)g_{4}(-\omega)-ig_{4}^{\ast}(-\omega)g_{3}(-\omega)\Big].
\end{eqnarray}

\begin{figure}[ht]
\centering
\includegraphics[width=2.8in]{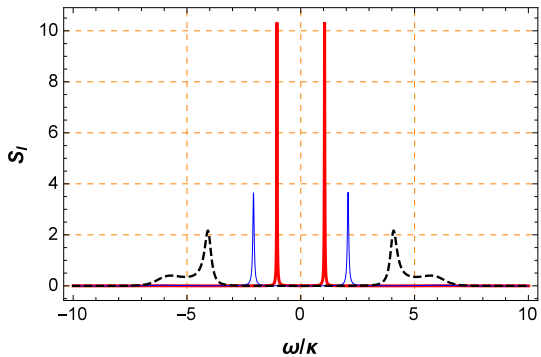}
\caption{
(Color online) The intensity power spectrum of the cavity output field versus the normalized frequency $ \omega/\kappa $ for three different values of the \textit{s}-wave scattering frequency $ \omega_{sw}=30\omega_{R} $ (red thick line), $ \omega_{sw}=60\omega_{R} $ (blue thin line), and $ \omega_{sw}=120\omega_{R} $ (black dashed line). Other parameters are the same as those in Fig.\ref{fig:fig3}.}
\label{fig:fig5}
\end{figure}

In Fig.\ref{fig:fig5} the intensity power spectrum of the cavity output field has been plotted versus the normalized frequency $ \omega/\kappa $ for three different values of the \textit{s}-wave scattering frequency $ \omega_{sw}=30\omega_{R} $ (red thick line), $ \omega_{sw}=60\omega_{R} $ (blue thin line), and $ \omega_{sw}=120\omega_{R} $ (black dashed line). The detuning has been fixed at $ \Delta_{c}=0.994\Delta_{0} $. 

Comparing Fig.\ref{fig:fig5} with Fig.\ref{fig:fig3}, one can see that while in the phase noise power spectrum both the central and the side peaks are completely observable for every value of the \textit{s}-wave scattering frequency, in the intensity power spectrum the side peaks are not observable except for very large values of $ \omega_{sw} $. Although the splitting of the two peaks for $ \omega_{sw}=30\omega_{R} $ and $ \omega_{sw}=60\omega_{R} $ are, respectively, $ \Delta\omega_{\pm}\approx 5.6\kappa $ and $ 4.2\kappa $, the side peaks are not observable because of their smallness in comparison to the central ones . However, for $ \omega_{sw}=120\omega $ the splitting reduces to $ \Delta\omega_{\pm}\approx 2.1\kappa $ but the side peaks are observable because of their largeness in comparison to the central ones. Therefore, the phase noise power spectrum is more suitable than the intensity spectrum for the measurement of the \textit{s}-wave scattering frequency.

\section{quadrature squeezing of the output field}
In this section we study the effect of atom-atom interaction on the squeezing properties of the transmitted field. The squeezing spectrum of the output field of cavity is given by
\begin{eqnarray}
S_{\varphi}(\omega)&=&\frac{1}{4\pi}\int d\omega^{\prime} e^{i(\omega+\omega^{\prime})t}\big\langle\delta X_{\varphi}^{(out)}(\omega)\delta X_{\varphi}^{(out)}(\omega^{\prime})\nonumber\\
&&+\delta X_{\varphi}^{(out)}(\omega^{\prime})\delta X_{\varphi}^{(out)}(\omega)\big\rangle.
\end{eqnarray}
where $ \delta X_{\varphi}^{(out)}(\omega)=e^{-i\varphi}\delta a_{out}(\omega)+e^{i\varphi}\delta a_{out}^{\dagger}(\omega) $ is the Fourier transform of the output quadrature, with $ \varphi $ as its externally controllable phase angle which is experimentally measurable in a homodyne detection scheme \cite{Loudon}. In this way, the squeezing spectrum is obtained as follows
\begin{eqnarray}\label{Saa}
S_{\varphi}(\omega)&=&e^{-2i\varphi}C_{aa}^{(out)}(\omega)+e^{2i\varphi}C_{a^{\dagger}a^{\dagger}}^{(out)}(\omega)\nonumber\\
&&+C_{aa^{\dagger}}^{(out)}(\omega)+C_{a^{\dagger}a}^{(out)}(\omega),
\end{eqnarray}
where
\begin{subequations}
\begin{eqnarray}
C_{aa}^{(out)}(\omega)&=&\frac{1}{4\pi} \int_{-\infty}^{+\infty}d\omega^{\prime} e^{i(\omega+\omega^{\prime})t}\langle \delta a_{out}(\omega)\delta a_{out}(\omega^{\prime})\nonumber\\
&&+\delta a_{out}(\omega^{\prime})\delta a_{out}(\omega)\rangle,\label{Caa}\\
C_{a^{\dagger}a}^{(out)}(\omega)&=&\frac{1}{4\pi} \int_{-\infty}^{+\infty}d\omega^{\prime} e^{i(\omega+\omega^{\prime})t}\langle \delta a^{\dagger}_{out}(\omega)\delta a_{out}(\omega^{\prime})\nonumber\\
&&+\delta a^{\dagger}_{out}(\omega^{\prime})\delta a_{out}(\omega)\rangle,\label{Cada}\\
C_{aa^{\dagger}}^{(out)}(\omega)&=&\frac{1}{4\pi} \int_{-\infty}^{+\infty}d\omega^{\prime} e^{i(\omega+\omega^{\prime})t}\langle \delta a_{out}(\omega)\delta a^{\dagger}_{out}(\omega^{\prime})\nonumber\\
&&+\delta a_{out}(\omega^{\prime})\delta a^{\dagger}_{out}(\omega)\rangle.\label{Caad}
\end{eqnarray}
\end{subequations}

The optimum quadrature squeezing $ S_{opt}(\omega) $ is defined by choosing $ \varphi $ such that $ dS_{\varphi}(\omega)/d\varphi=0 $ which yields
\begin{equation}
e^{2i\varphi_{opt}}=\pm\frac{C_{aa}^{(out)}(\omega)}{\vert C_{aa}^{(out)}(\omega)\vert}.
\end{equation}
We choose the minus solution in order to minimize the spectrum function. By substituting this solution into Eq.(\ref{Saa}) the optimized squeezing spectrum will be obtained as follows
\begin{equation}\label{Sopt}
S_{opt}(\omega)=-2\vert C_{aa}^{(out)}(\omega)\vert+C_{aa^{\dagger}}^{(out)}(\omega)+C_{a^{\dagger}a}^{(out)}(\omega).
\end{equation}

Using the input-output relations [Eqs.(\ref{in-out1}, \ref{in-out2})], Eqs.(\ref{Caa},\ref{Cada},\ref{Caad}) can be written as
\begin{subequations}
\begin{eqnarray}
C_{aa}^{(out)}(\omega)&=&2\kappa C_{aa}(\omega),\\
C_{a^{\dagger}a}^{(out)}(\omega)&=&2\kappa C_{a^{\dagger}a}(\omega),\\
C_{aa^{\dagger}}^{(out)}(\omega)&=&2\kappa C_{aa^{\dagger}}(\omega)+1,
\end{eqnarray}
\end{subequations}
where the expressions for $ C_{aa}(\omega) $, $ C_{a^{\dagger}a}(\omega) $ and $ C_{aa^{\dagger}}(\omega) $ can be obtained from Eqs.(\ref{Caa},\ref{Cada},\ref{Caad}) by replacing $ a_{out} $ and $ a_{out}^{\dagger} $ with $ a $ and $ a^{\dagger} $, respectively. In this way, the optimum quadrature squeezing [Eq.(\ref{Sopt})] can be written as
\begin{equation}
S_{opt}(\omega)=-4\kappa\vert C_{aa}(\omega)\vert+2\kappa C_{aa^{\dagger}}(\omega)+2\kappa C_{a^{\dagger}a}(\omega)+1.
\end{equation}

Using Eqs.(\ref{a(w)1},\ref{a(w)2}) the optimum quadrature squeezing can be obtained as a function of the coefficients $ g_{i}(\omega) $ and the susceptibility $ \chi(\omega) $ . The squeezing occurs when $ S_{opt}(\omega)<1 $. Besides, $ S_{opt}(\omega)=1 $ corresponds to the spectrum of fluctuations for a vacuum or coherent field. 

\begin{figure}[ht]
\centering
\includegraphics[width=2.8in]{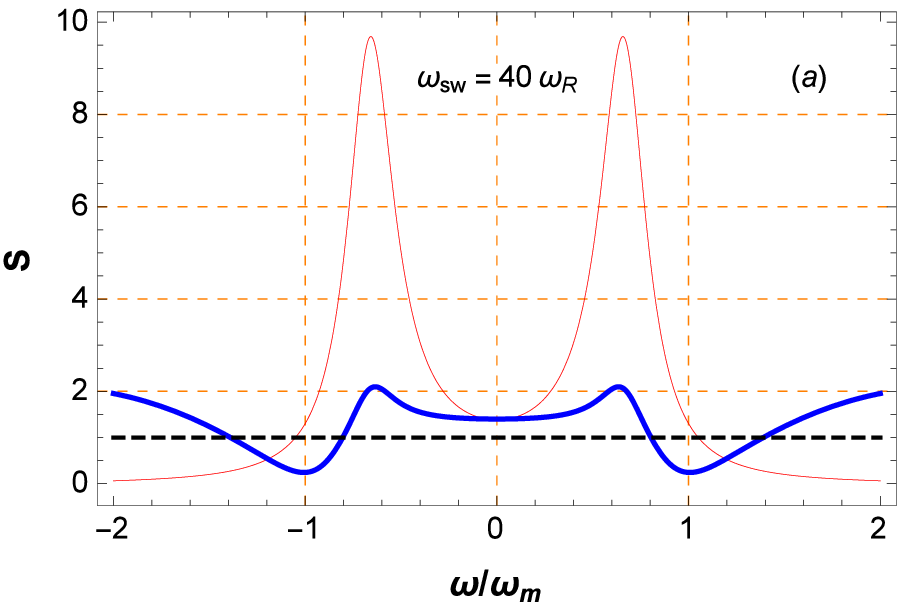}
\includegraphics[width=2.8in]{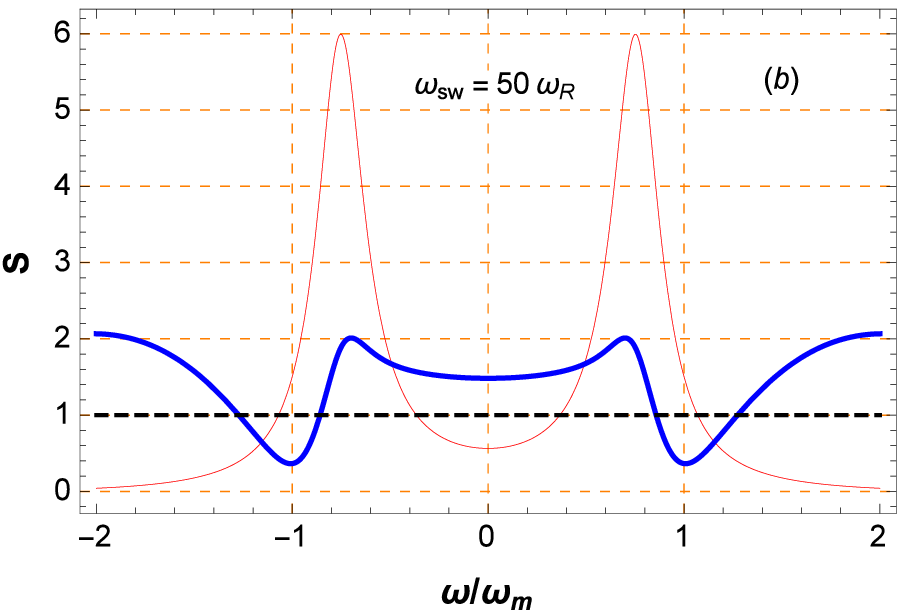}
\includegraphics[width=2.8in]{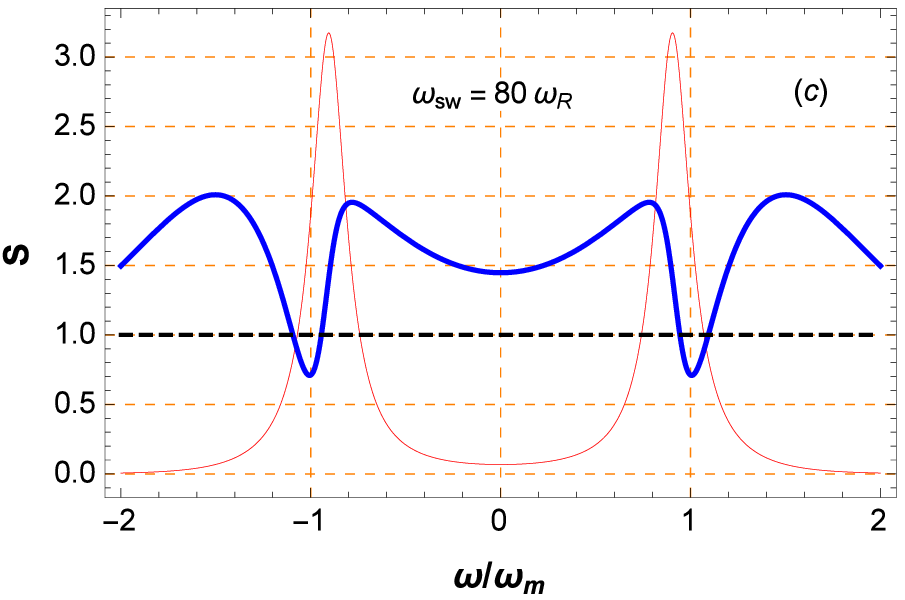}
\caption{
(Color online) The intensity power spectrum (red thin line) and the squeezing spectrum (blue thick line) of the output field  of the cavity versus the normalized frequency $ \omega/\omega_{m} $ for $ \delta_{c}=\omega_{m} $ and $ \kappa=74\omega_{m} $ and for three different values of the \textit{s}-wave scattering frequency (a) $ \omega_{sw}=40\omega_{R} $, (b) $ \omega_{sw}=50\omega_{R} $, and (c) $ \omega_{sw}=80\omega_{R} $. Other parameters are the same as those in Fig.\ref{fig:fig2}.}
\label{fig:fig6}
\end{figure}

In order to see the effect of atomic collisions on the squeezing of the transmitted field of the cavity, in Fig.\ref{fig:fig6} we have plotted the quadrature squeezing spectrum (blue thick line) as well as the intensity spectrum (red thin line) of the output field  of the cavity versus the normalized frequency $ \omega/\omega_{m} $ for $ \delta_{c}=\omega_{m} $ and $ \kappa=74\omega_{m} $. The spectra have been plotted for three different values of the \textit{s}-wave scattering frequency: $ \omega_{sw}=40\omega_{R} $ [Fig.\ref{fig:fig6}(a)], $ \omega_{sw}=50\omega_{R} $ [Fig.\ref{fig:fig6}(b)], and $ \omega_{sw}=80\omega_{R} $ [Fig.\ref{fig:fig6}(c)]. The black dashed line corresponds to the spectrum of fluctuations for a vacuum or coherent field where $ S_{opt}(\omega)=1 $.

As is seen from Fig.\ref{fig:fig6}, for each value of $ \omega_{sw} $ the squeezing spectrum exhibits two dips centered at $ \vert\omega\vert=\omega_{m} $ below the coherent level [$ S_{opt}(\omega)=1 $] while the peaks of the intensity spectrum occur a little bit below $ \omega_{m} $. The positions of these intensity peaks get nearer to $ \omega_{m} $ by increasing the \textit{s}-wave scattering frequency. 

Generally, the atom-atom interaction shifts the resonance frequency of the cavity \cite{dalafi1} which leads to a reduction in the intensity of the transmitted field. That is why the peaks of $ S_{P}(\omega) $ and $ S_{I}(\omega) $ reduce by increasing the \textit{s}-wave scattering frequency. This effect may also reduce a little bit the amount of the squeezing of the output field. Nevertheless, increasing the atom-atom interaction cause the peaks of the intensity spectrum get nearer to the dips of the squeezing spectrum, as is observable in Fig.\ref{fig:fig6}. In other words, although the amount of the output squeezing for lower values of $ \omega_{sw} $ is somehow more than that of higher values, but this squeezing occurs where the intensity of the output field is very weak while for higher values of $ \omega_{sw} $ the maximum amount of squeezing occur where the output intensity is considerable. In this way one can assert that atomic collisions have positive effect on the production of squeezed light.

\section{Summary and Conclusion}\label{secCon}
In this paper, we have studied an interacting one-dimensional BEC inside an optical cavity which is driven through one of the fixed end mirrors. Under the Bogoliubov approximation and when the number of photons inside the cavity is not too large, the atomic field operator can be considered as a single-mode quantum field which is coupled to the radiation pressure of the intracavity field. In this way, the system behaves like an optomechanical system with an extra nonlinear term corresponding to the atom-atom interaction.

We have shown that one of the best ways of tracing the effect of atomic interaction is to study the phase noise power spectrum of the output field of the cavity. For this purpose, we studied the phase noise and intensity power spectra as well as the quadrature squeezing of the transmitted field of the optical cavity. The results reveal that the effect of atomic collisions is manifested as a change in the splitting between the normal modes of the system. However, this behavior can be observed much more clearly in the phase noise spectrum than that the intensity power spectrum. We have also derived a one-to-one correspondence between the amount of splitting between the normal modes of the transmitted field of the cavity and the \textit{s}-wave scattering frequency of the atomic collisions. Therefore, by measuring the frequency splitting of the two peaks of the phase noise power spectrum which is experimentally feasible by the homodyne measurement of the light reflected by the cavity, one can estimate the value of \textit{s}-wave scattering frequency of the atoms.

We have also examined the effect of atomic interactions on the squeezing behavior of the transmitted field of the cavity. Based on our results, although for lower values of the \textit{s}-wave scattering frequency the degree of squeezing is somehow larger, this squeezing occurs at the frequency in which the output intensity power spectrum is nearly zero (the frequency of the BEC mechanical mode). By increasing the \textit{s}-wave scattering frequency the peak position of the intensity power spectrum gets nearer to the frequency of the BEC mechanical mode where the transmitted light is squeezed. It means that in order to achieve output light squeezing with considerable intensity one needs to increase the strength of atomic collisions.

\section*{Acknowledgement}
A.D wishes to thank the Laser and Plasma Research Institute of Shahid Beheshti University for its support.

\bibliographystyle{apsrev4-1}

\end{document}